\documentclass[journal]{rmaa}

\usepackage{paralist}
\usepackage{graphicx}

\usepackage{psfrag,color}

\usepackage[latin1]{inputenc}

\ReceivedDate{June 5 2017}
\AcceptedDate{October 5 2017}
\title{Emission line galaxies behind the planetary nebula IC~5148: \\Potential for a serendipity survey with archival data.}
\author{
S. Kimeswenger\altaffilmark{1,2}
D. Barria\altaffilmark{1}
W. Kausch\altaffilmark{2}
D.~S. Goldman\altaffilmark{3}}
\altaffiltext{1}{Instituto de Astronom{\'i}a, UCN, Antofagasta, Chile.}
\altaffiltext{2}{Institut f. Astro- und Teilchenphys., Innsbruck, Austria.}
\altaffiltext{3}{Astrodon Imaging, Roseville, USA.}
\shortauthor{Kimeswenger et al.}
\shorttitle{ELGs behind the PN IC~5148: Potential for a serendipity survey.}

\fulladdresses{
\item Stefan Kimeswenger and Daniela Barria: Instituto de Astronom{\'i}a, Universidad Cat{\'o}lica del Norte, Av. Angamos 0610, Antofagasta, Chile.
\item Stefan Kimeswenger and Wolfgang Kausch: Institut f{\"u}r Astro- und Teilchenphysik, Leopold--Franzens Universit{\"a}t Innsbruck, Technikerstr. 25, 6020 Innsbruck, Austria.
\item Don S. Goldman: Astrodon Imaging, 407 Tyrolean Court. Roseville, California, 95661, USA.}
\listofauthors{S. Kimeswenger, D. Barria, W. Kausch, \& D.S. Goldman}
\indexauthor{Kimeswenger, S.}
\indexauthor{Barria, D.}
\indexauthor{Kausch, W.}
\indexauthor{Goldman, D.S.}

\abstract{During the start of a survey program using FORS2 long slit spectroscopy on planetary nebulae (PN) and their haloes, we serendipitously discovered six background emission line galaxies (ELG) with redshifts of $z =$ 0.2057, 0.3137, 0.37281, 0.4939, 0.7424 and 0.8668. Thus they clearly do not belong to a common cluster structure. We derived the major physical properties of the targets. Since the used long slit covers a sky area of only 570 arcsec$^2$ ($= 4.3\times10^{-5}$ square degrees), we discuss further potential of serendipitous discoveries in archival data, beside the deep systematic work of the ongoing and upcoming big surveys. We conclude that archival data provide a decent potential for extending the overall data on ELGs without any selection bias.}

\resumen{Durante el comienzo de un programa de inspecci{\'o}n usando espectroscopia de rendija larga con FORS2 en PNe y sus halos, descubrimos por casualidad seis galaxias de fondo de emisi{\'o}n (ELGs) con valores de corrimiento al rojo de $z =$ 0.2057, 0.3137, 0.37281, 0.4939, 0.7424 y 0.8668, respectivamente. Claramente estas galaxias no pertenecen a un c{\'u}mulo en com{\'u}n. Hemos derivado las principales propiedades f{\'i}sicas de estos objetos. Considerando que se ha usado una rendija larga que cubre un {\'a}rea del cielo de tan solo 570 arcsec$^2$ ($= 4.3\times10^{-5}$ grados cuadrados), discutimos el futuro potencial en el uso de datos de archivo en el descubrimiento por casualidad de estos objetos, mas all{\'a}~ de los extensos programas de inspeccion actualmente en desarrollo y los futuros. Hemos concluido que estos datos de archivo representan un verdadero potencial para extender el total de datos de ELGs sin efectos de selecci{\'o}n.}

\addkeyword{galaxies: active}
\addkeyword{galaxies: distances and redshifts}
\addkeyword{methods: observational}
\addkeyword{techniques: spectroscopic}
\addkeyword{surveys}

\begin{document}
\maketitle


%
%
%
%
%
%


\section{Introduction}

Emission line galaxies (ELGs) are frequently found in the universe. They are important to study environments of star formation and the more active environments in active galactic nuclei. New massive multi object (MOS) spectrograph surveys on ELGs (e.g. eBOSS on SDSS-IV) observe dedicated fields like the polar caps \citep[see e.g.][and references therein]{eBOSS} and cover a few hundred thousand targets soon. They use selections by photometric characteristics of ELGs to position the slits. Tests have shown that this will introduce only a marginal selection bias due to the size of the samples. The pilot survey already covered 9000 spectra \citep{eBOSSpilot}. A very recent survey of the Multi Unit Spectroscopic Explorer (MUSE) consortium during guaranteed time \citep{MUSE} aiming in detection of ELGs will cover completely the area with this new instrument. No photometric selection is required as no  MOS slits have to be positioned. It finally will cover 120 arcmin$^2$ of the CANDELS/Deep area of the Chandra Deep Field South.

On the other hand serendipitous surveys are normally only carried out during space missions and analysis of space mission data, e.g.  {\sl The Cambridge-Cambridge ROSAT Serendipity Survey} \citep{CCRSS}, {\sl The ISO Serendipity Survey} \citep{ISOS}, {\sl The ASCA Hard Serendipitous Survey} \citep{ASCAS}, {\sl The XMM-Newton Serendipitous Survey} \citep{XMMS}, {\sl The Swift XRT serendipitous deep survey} \citep{SwiftS} and {\sl The NuSTAR Serendipitous Survey} \citep{NuSTAR2}. Future plans also include next generation space telescopes like the James Webb Space Telescope incorporating the Medium Resolution Spectrometer \citep{JWST}.

They show that in addition to dedicated surveys, existing archival data provide a huge amount of data for serendipitous detections. This is, as the observations were taken for different other purposes originally, without any selection bias. In this paper we report on the discovery of six ELGs in FORS2 long-slit spectra taken within the framework of a detailed investigation of a recently discovered thin halo around the planetary nebula (PN) IC 5148 and discuss the potential of further serendipitous discoveries by a survey of the whole archive of this instrument.

\section{Data}

The spectra were taken with the FOcal Reducer and low dispersion Spectrograph 2 (FORS2) \citep{fors} mounted on the Cassegrain focus of ESO VLT UT1 (Antu) in the nights October 5$^{\rm th}$, 2016 from 3:15 to 4:03\,UT, October 6$^{\rm th}$, 2016 from 0:15 to 0:30\,UT and October 10$^{\rm th}$, 2016 from 1:17 to 1:45\,UT in service mode.
In total 7 spectra were obtained, 3 with a position angle 30$^{\rm o}$~and 4 with a position angle 150$^{\rm o}$~(from N over E) on the sky trough the center of the PN IC\,5148. The slit position was selected to cover some features in wide halo of the PN. We used the long-slit mode of FORS2 since this enabled us to cover the entire halo  using the full slit length of 6\farcm8 with the standard collimator (slit width = 0\farcs7). All spectra were taken with 14 minutes exposure time.
We used GRISM 300V and the GG435 order separation filter, covering a wavelength range from 455 to 889\,nm. The MIT/LL CCD mosaic and the standard focal reducer collimator result in a 0\farcs2518 pixel$^{-1}$ spatial resolution. This setup leads to a final wavelength dispersion of 0.33\,nm\,pixel$^{-1}$. The night sky lines were measured with a resolution $R = \Delta\lambda/\lambda$ of 200 and $R = 360$ at the blue end and the red end of the spectrum, respectively. We derived a FWHM of the stellar sources along the slit of about 1\farcs17 at the blue end and 0\farcs95 at the red part of the spectrum. This corresponds well to the reported DIMM seeing of the ESO meteo monitor of 1\farcs1\,@500\,nm.
The data were reduced incorporating the standard calibration mode using the ESO FORS pipeline v5.3.11 \citep{pipeline}. 
The resulting flux calibration was compared with the expected continuum flux of the central star of the PN finding differences less than 2.5\%. We used the software package {\sl molecfit} \citep{molecfit1,molecfit2} to create a model of the telluric absorption lines by applying it on the high S/N central star spectrum. As the PN and its extended halo covered the whole slit length we used the software {\sl skycorr} \citep{skycorr} to correct for the sky emission lines. This approach enabled us to derive a post correction of the wavelength calibration, too. Finally we averaged the spectra for each sky direction to achieve the final spectra of the faint sources used for this study.

\section{Results and Discussion}
\begin{figure}[!ht]
   \centering
   \includegraphics[width=78mm]{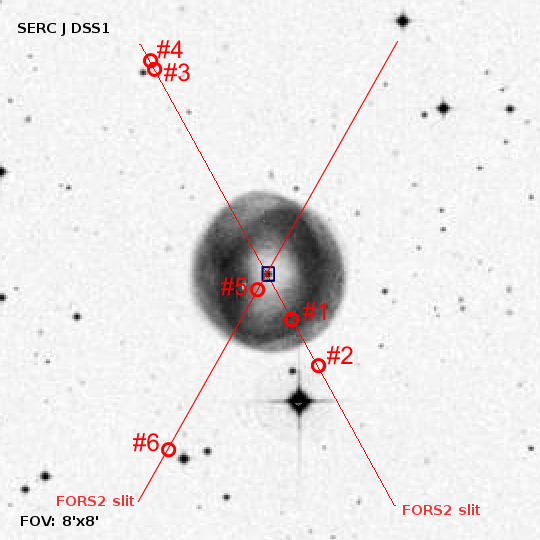}\\
   \smallskip
   \includegraphics[width=78mm]{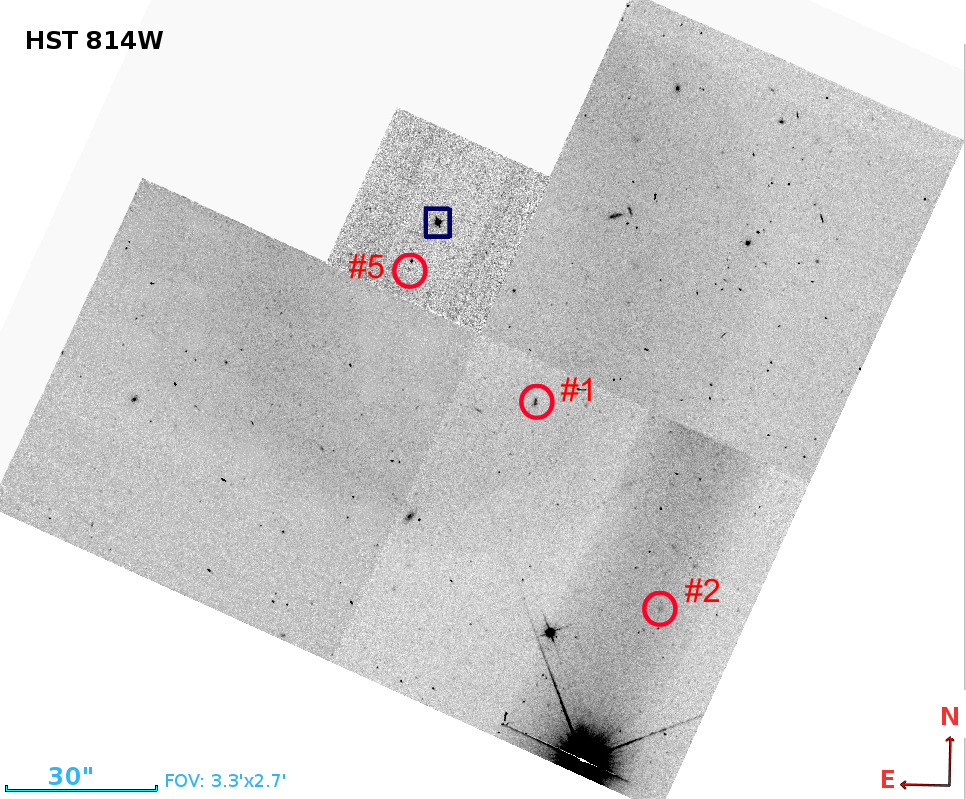}
      \caption{Field around IC5148. The upper panel is based on the blue sky survey blue SERC DSS J$_{\rm ph}$. The emission line galaxies are marked as circles. The lower panel covers the center of the field observed with the HST WFPC2 with a very red optical filter F814W. It shows three of the objects and a rich galaxy field behind the PN. The black square marks the central star of the PN. }
         \label{finding}
   \end{figure}

Fig.~\ref{finding} shows the field with the PN, and the slit including the positions of the ELGs. Since the targets are hardly visible on the Digitized Sky Survey (DSS) frame, we also show the 200 seconds HST WFPC2 exposure taken with the F814W filter, which covers the southern part of the PN and covers 3 of our targets.

Due to the difference redshift, clearly they do not belong to a common cluster structure. We derived the coordinates of target \#1, \#2 and \#5 by means of the HST field, being therefore more accurate than the coordinates of targets \#3, \#4 and \#6, which were determined using the location of the emission lines in the slit relative to the central star of the PN.

\noindent To calculate the distances, luminosities and redshift based color corrections, we used the online cosmology calculator by \citet{coscalc1}, linked in the NASA/IPAC Extragalactic Database (NED). We further used the same cosmology ($H_0 = 73.0$\,\,km\,\,s$^{-1}$\,\,Mpc$^{-1}$, $\Omega_{\rm Matter} = 0.27$ and $\Omega_{\rm Vacuum} = 0.73$) as included in NED. The applied extinction corrected was the method of \citet{extinct_old} in the form recalibrated by \citet{extinction}.

\subsection{Galaxy \#1  ($z=0.2055$): J215933.45$-$392343.9}
\begin{figure}[!ht]
   \centering
   \includegraphics[width=78mm]{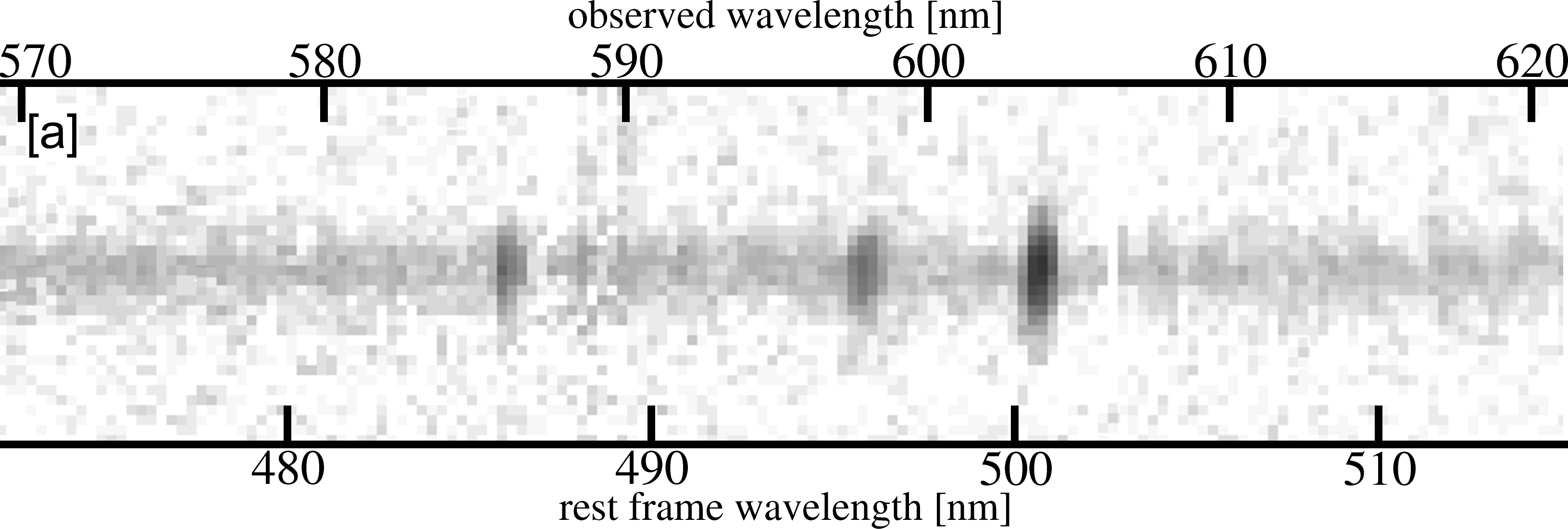}\\
   \smallskip
   \includegraphics[width=78mm]{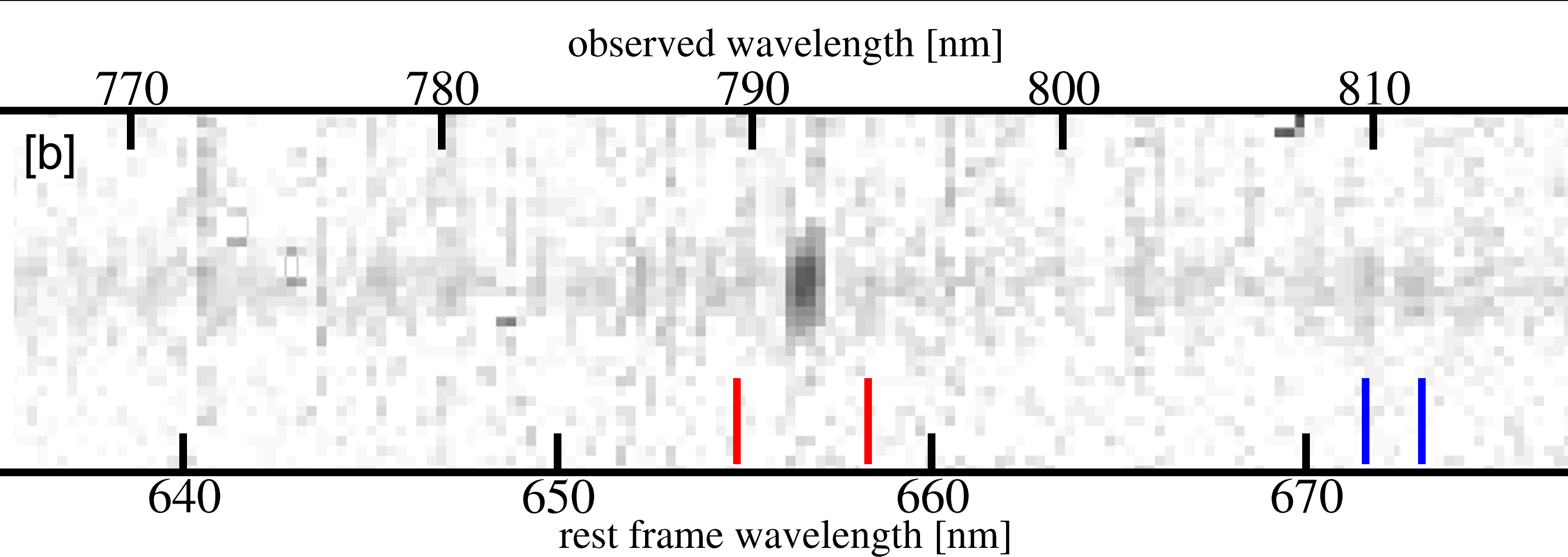}\\
   \smallskip
   \includegraphics[width=78mm]{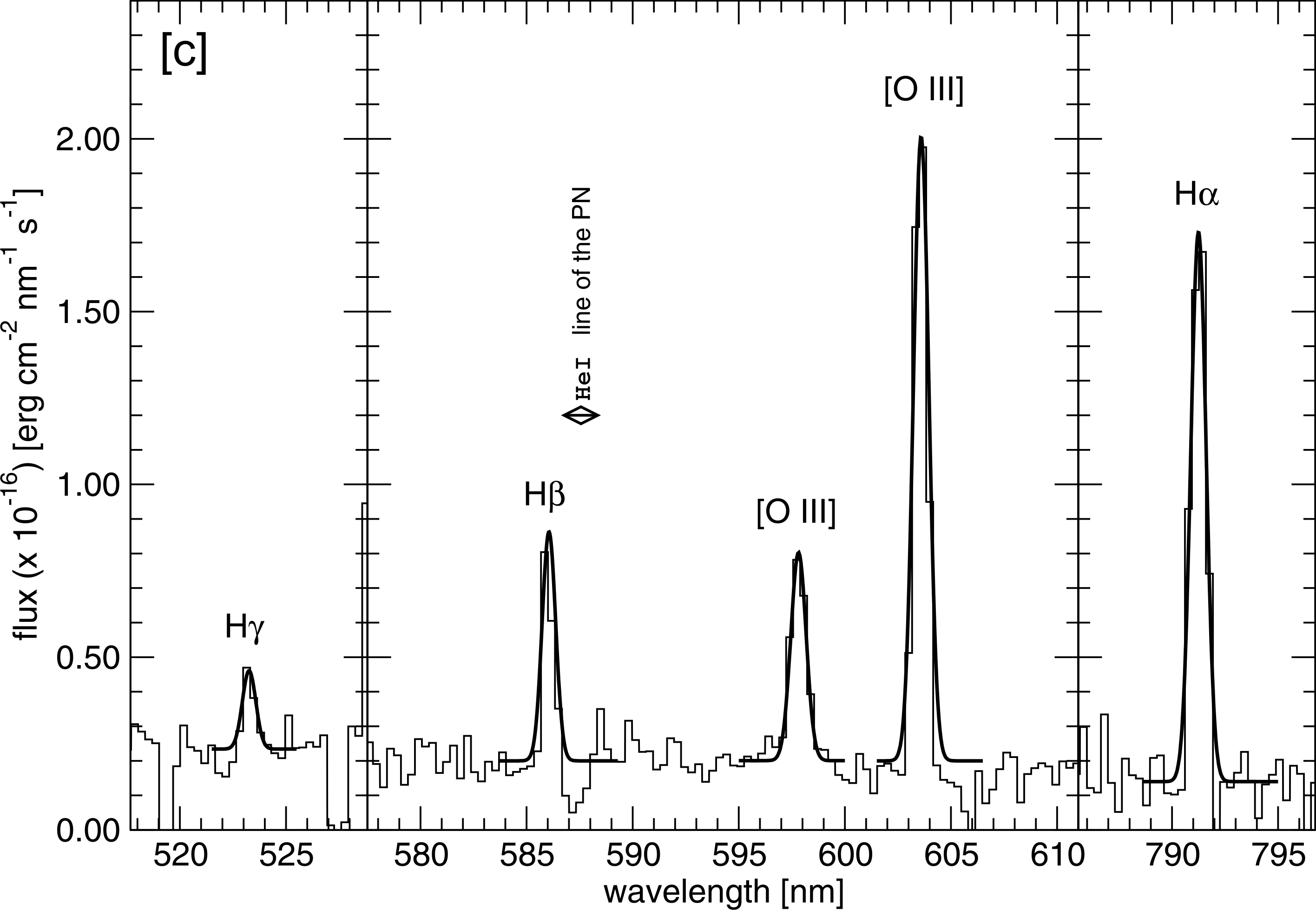}
      \caption{Galaxy \#1: The 2D spectrum around \ion{H}{$\beta$}+[\ion{O}{iii}] (panel [a]) and around \ion{H}{$\alpha$} (panel [b]) and the integrated spectrum (panel [c]). The red ticks correspond to the redshifted wavelengths of the [\ion{N}{ii}] lines at 654.8 and 658.4\,nm. The blue tics show the weak detection of the [\ion{S}{ii}]~671.6+673.1\,nm pair.
      Each panel covers 50\,nm in wavelength and 10\arcsec~in spatial direction. We have chosen equal grey--scale for all those figures for easier comparison. The gray scale range is [0,$2\times10^{-19}$\,erg\,cm$^{-2}$\,nm$^{-2}$\,s$^{-1}$]. Panel [c] shows the integrated spectrum of the galaxy and the best gaussian fits as overlay. In case of \ion{H}{$\beta$} a strong contamination by the subtracted \ion{He}{i} line of the PN exists (see text for more details).}
         \label{G1_image}
   \end{figure}

This galaxy is completely covered by the main rim of the PN. The spectral regions around the emission lines are given in Fig.~\ref{G1_image}. The nitrogen lines are only upper limits and the sulphur lines marginally detected since they are near the noise level. From the integrated spectrum of all 3 frames we obtained individual redshifts for the lines (c.f. Table~\ref{G1_tab}).

\noindent The \ion{H}{$\beta$} line shows a slightly smaller profile and a slightly lower redshift then all other lines. The profile even was somewhat smaller than that measured on night sky lines. Since we removed a very strong \ion{He}{i} line of the PN at its red edge we assume a contamination and use only the blue part for a final fit of the intensity. Furthermore this line was not included to the total redshift calculation. The redshifts, line intensities, target coordinates, the calculated distance modulus, and the line luminosities are summarized in Table~\ref{G1_tab}. The Balmer series resemble a \ion{H}{$\alpha$}/\ion{H}{$\beta$} line ratio, which is nonphysical for electron temperatures above a few thousands of degrees \citet{OF06}. Moreover, the \ion{H}{$\alpha$}/\ion{H}{$\gamma$} is, within the expected errors of a few percent in line fluxes as expected. We thus have to assume that the contamination of \ion{H}{$\beta$} by the PN line mentioned above caused a flux error of up to 15\% for this line.
The continuum in the spectrum resembles a fairly blue object, although our blue cutoff at 450nm does not show the region around the Balmer jump at the given redshift.
\begin{table}[!ht]
\caption{Summary of the observed parameters of galaxy \#1.}
\label{G1_tab}
{\centering
\begin{tabular}{l c c c c}
\toprule
 position$\!\!\!\!\!\!$& \multicolumn{4}{l}{\qquad$\alpha_{\rm J2000} = \,\,\,\,21^{\rm h}59^{\rm h}33\fs45$}\\
\phantom{position}$\!\!\!\!\!\!$& \multicolumn{4}{l}{\qquad$\delta_{\rm J2000}\,=-39^{\rm o}23\arcmin43\farcs9$ }
\smallskip\\
ion  & $\lambda_{rest}$ & $\lambda_{obs}$ & $z$ & flux \\
     & nm & nm &   & $[^*]$\\
\ion{H}{$\gamma$} & 434.047 & 523.249 & 0.2055 & \phantom{1}2.0 \\
\ion{H}{$\beta$}  & 486.135 & 585.968 & \,\,\,0.2054$^{\rm a}$ & \phantom{1}6.1\\
$[$\ion{O}{iii}$]$    & 495.891 & 597.800 & 0.2055 & \phantom{1}5.6 \\
                  & 500.684 & 603.557 & 0.2055 & 16.7 \\
\ion{H}{$\alpha$} & 656.279 & 791.254 & 0.2057 & 14.4 \\
$[$\ion{S}{ii}$]$ & 671.644 & \,\,\,809.694$^{\rm b}$ & & $\!\!\!\!\leq\,$ 0.7\\
                  & 673.082 & \,\,\,811.427$^{\rm b}$ & & $\!\!\!\!\leq\,$ 0.7\smallskip\\
redshift$\!\!\!\!\!\!$ & \multicolumn{4}{l}{\qquad$z_{\rm tot}=0.2055\pm0.0001$}
\smallskip\\
luminosity$\!\!\!\!\!\!\!\!\!\!\!\!\!$ & \multicolumn{4}{l}{\qquad$L_{{\rm H}\alpha}\,=1.6\,\times\,10^{41} {\rm erg}\,{\rm cm}^{-2}\,{\rm s}^{-1}$}\\
 & \multicolumn{4}{l}{\qquad$L_{\ion{O}{iii}}=1.9\,\times\,10^{41} {\rm erg}\,{\rm cm}^{-2}\,{\rm s}^{-1}$}\\
\bottomrule
\end{tabular}}
$[^*]$ {\small flux unit: $\times 10^{-16}$erg\,cm$^{-2}$\,s$^{-1}$)}\protect{\newline}
$^{\rm a}\,\,\,\,${\small Not included to calculate value of $z_{\rm tot}$ due to}
\protect{\newline}\phantom{$^{\rm a}\,\,\,\,$}{\small PN contamination.}\protect{\newline}
$^{\rm b}\,\,\,\,${\small Not measured but calculate by $z_{\rm tot}$.}
\end{table}


\noindent


\subsection{Galaxy \#2 ($z=0.3720$): J215931.28$-$392425.4}
\begin{figure}[!ht]
   \centering
   \includegraphics[width=78mm]{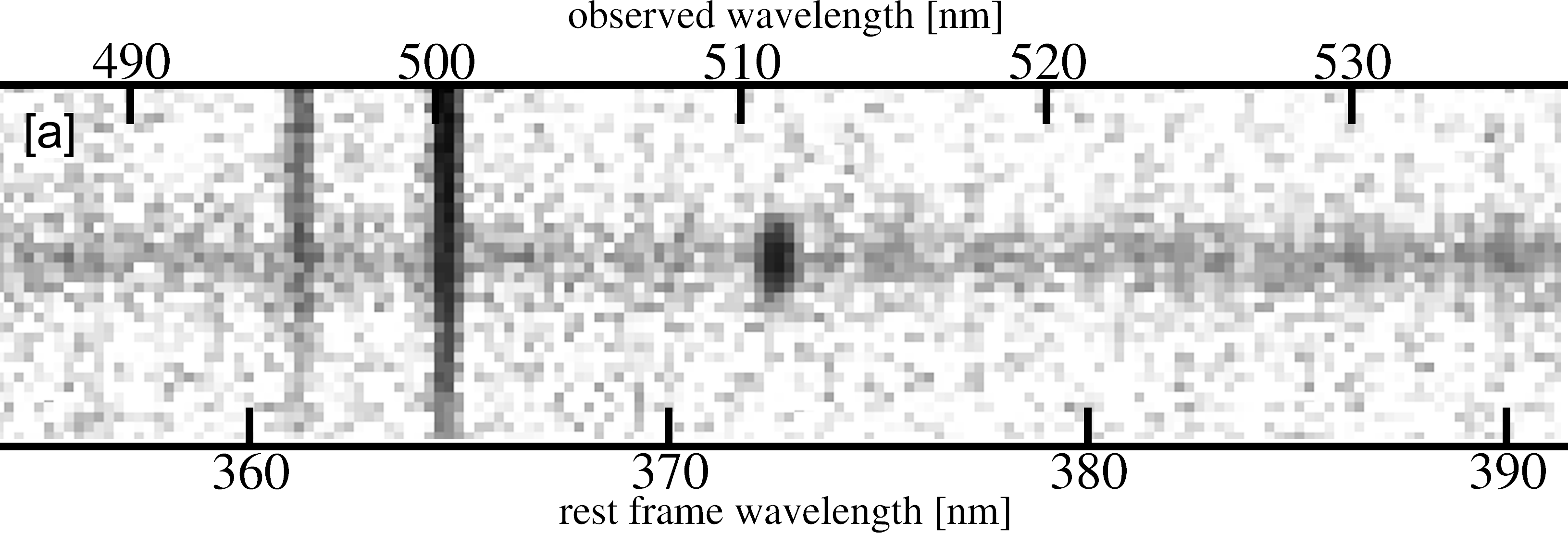}\\
   \smallskip
   \includegraphics[width=78mm]{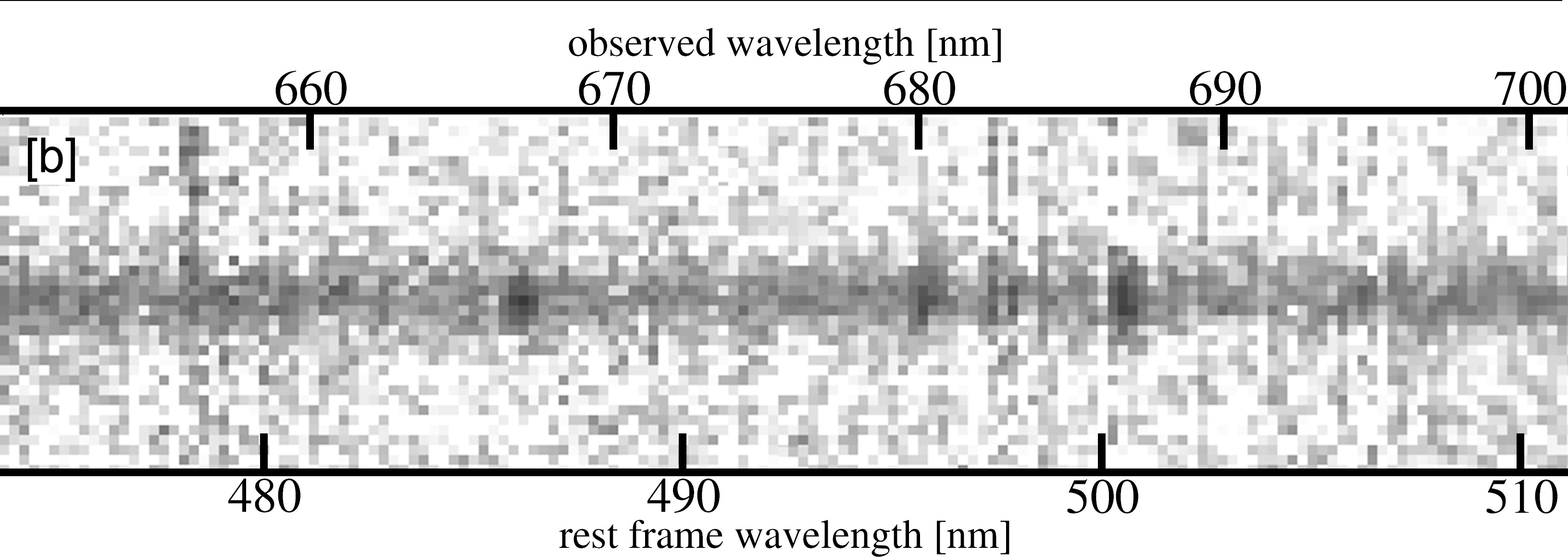}\\
   \smallskip
   \includegraphics[width=78mm]{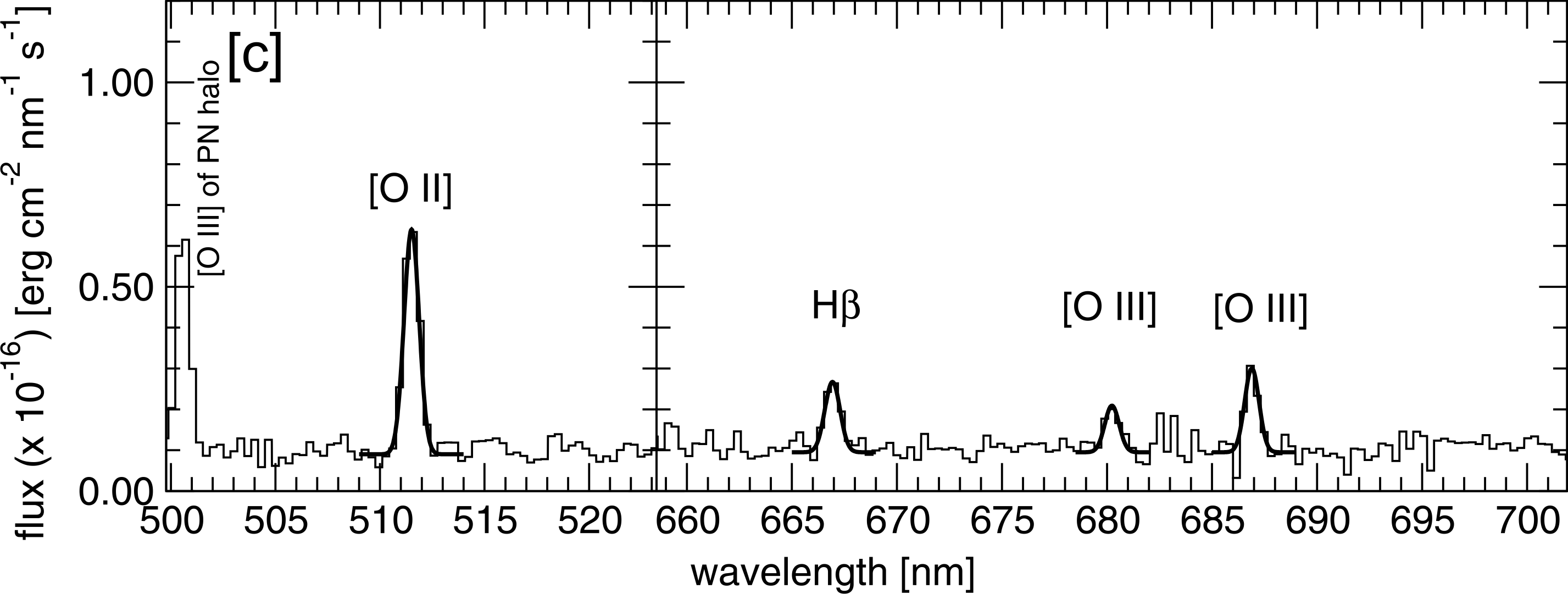}
      \caption{Galaxy \#2: The spectrum around [\ion{O}{ii}] (panel [a]) and \ion{H}{$\beta$}+[\ion{O}{iii}] (panel [b]) using the same intensity scale and size as Fig.~\ref{G1_image}. The
      lines crossing the whole upper panel are the \ion{O}{iii} halo emission lines of the PN. The integrated spectrun is given in panel [c] together with the best line fits in overlay.}
         \label{G2_image}
   \end{figure}

This galaxy appears very faint on the HST/F814W image. There is a whole group of such low surface brightness objects on the image visible within a radius of about 0.5\arcmin. However, the emission lines are in good contrast to the sky background and the noise (see Fig.~\ref{G2_image}, Table~\ref{G2_tab})).

The galaxy is, although dust extinction would affect the oxygen line luminosity more than the hydrogen lines, exceptionally bright in its \ion{O}{ii} emission (Table~\ref{G2_tab}).

\begin{table}[!ht]
\caption{Summary of the observed parameters of galaxy \#2.}
\label{G2_tab}
{\centering
\begin{tabular}{l c c c c}
\toprule
position$\!\!\!\!\!\!$ & \multicolumn{4}{l}{\qquad$\alpha_{\rm J2000} = \,\,\,\,21^{\rm h}59^{\rm h}31\fs28$}\\
\phantom{position}$\!\!\!\!\!\!$& \multicolumn{4}{l}{\qquad$\delta_{\rm J2000}\,=-39^{\rm o}24\arcmin24\farcs4$ }
\smallskip\\
ion  & $\lambda_{rest}$ & $\lambda_{obs}$ & $z$ & flux  \\
     & nm & nm &   & [*] \\
$[$\ion{O}{ii}$]$ & $\,\,\,$372.742$^{\rm a}$ & 511.512 & 0.3723 & \phantom{1}5.1 \\
\ion{H}{$\beta$}  & 486.135 & 666.938 & 0.3719 & \phantom{1}1.6\\
$[$\ion{O}{iii}$]$    & 495.891 & 680.236 & 0.3717 & \phantom{1}0.7 \\
                  & 500.684 & 686.896 & 0.3719 & \phantom{1}1.9\smallskip\\
redshift$\!\!\!\!\!\!$ & \multicolumn{4}{l}{\qquad$z_{\rm tot}=0.3720\pm0.0002$}
\smallskip\\
luminosity$\!\!\!\!\!\!\!\!\!\!\!\!\!$ & \multicolumn{4}{l}{\qquad$L_{{\rm H}\beta}\,=7.1\,\times\,10^{40} {\rm erg}\,{\rm cm}^{-2}\,{\rm s}^{-1}$}\\
 & \multicolumn{4}{l}{\qquad$L_{\ion{O}{iii}}=8.4\,\times\,10^{40} {\rm erg}\,{\rm cm}^{-2}\,{\rm s}^{-1}$}\\
 & \multicolumn{4}{l}{\qquad$L_{\ion{O}{ii}}\,=2.3\,\times\,10^{41} {\rm erg}\,{\rm cm}^{-2}\,{\rm s}^{-1}$}

\smallskip\\
\bottomrule
\end{tabular}}
$^{\rm a}\,\,\,\,${\small Calculated as mean of 372.603 and 372.881\,nm}\protect{\newline}
\phantom{$^{\rm a}\,\,\,\,$}{\small assuming about equal line strengths in the blend.}
\end{table}


\subsection{Galaxy \#3 ($z=0.4937$): J215944.0$-$392015}
\begin{figure}[!ht]
   \centering
   \includegraphics[width=78mm]{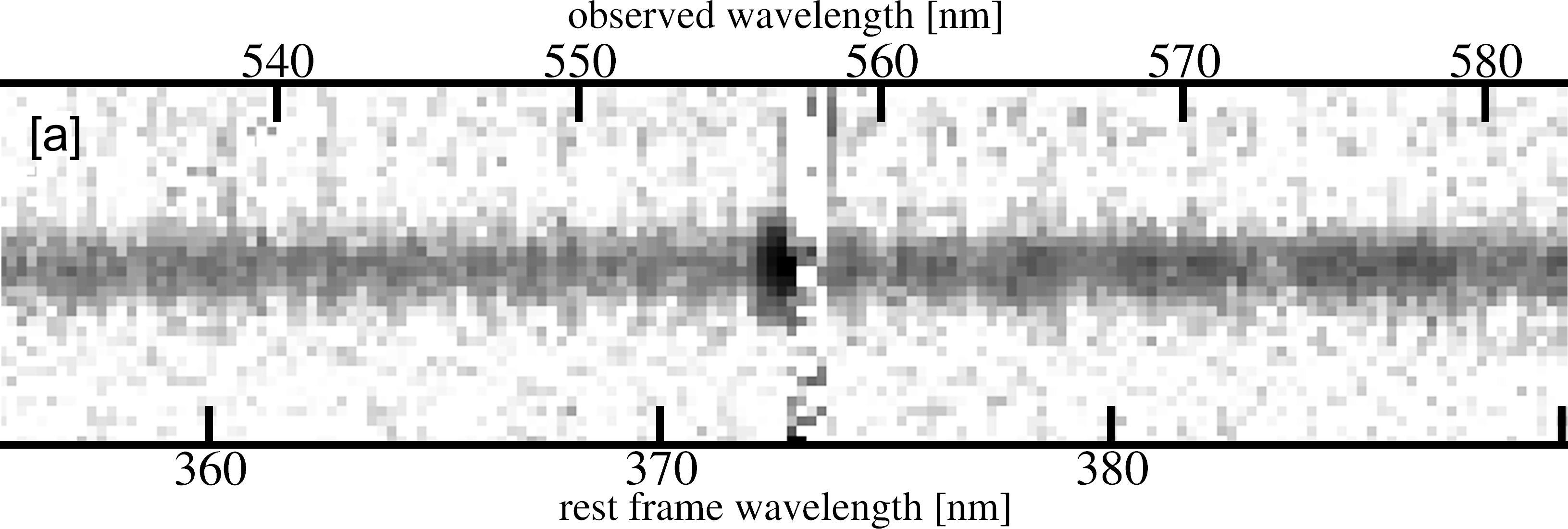}\\
   \smallskip
   \includegraphics[width=78mm]{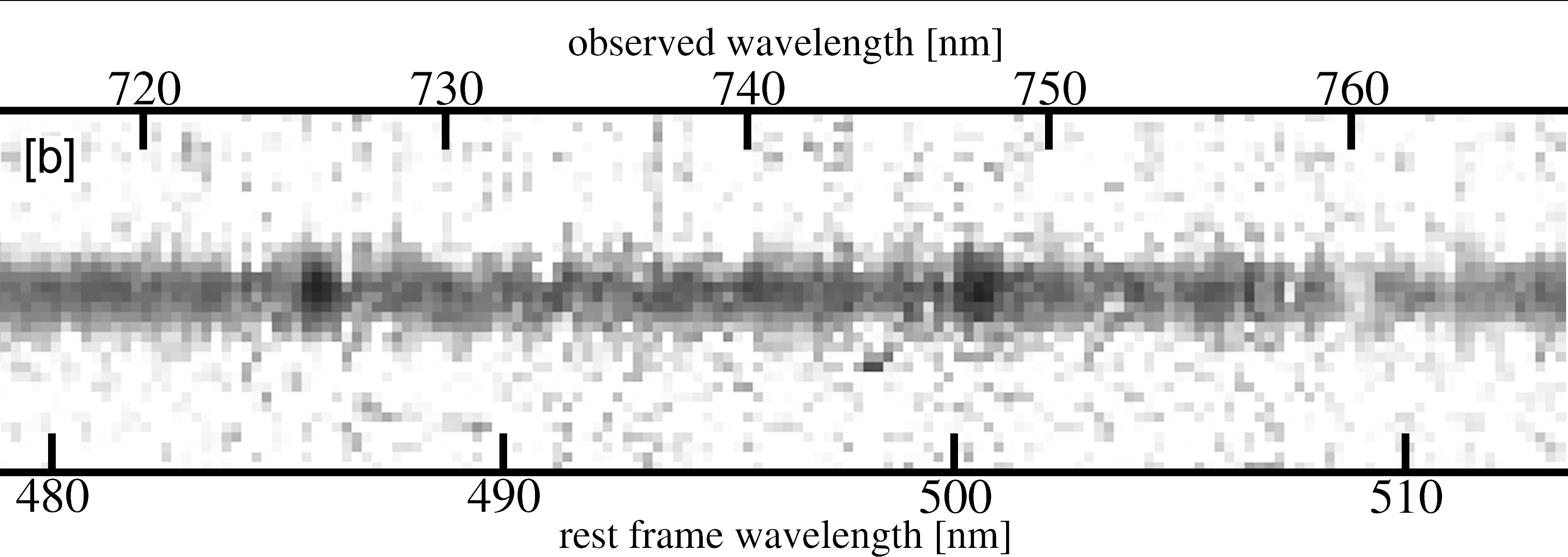}\\
   \smallskip
   \includegraphics[width=78mm]{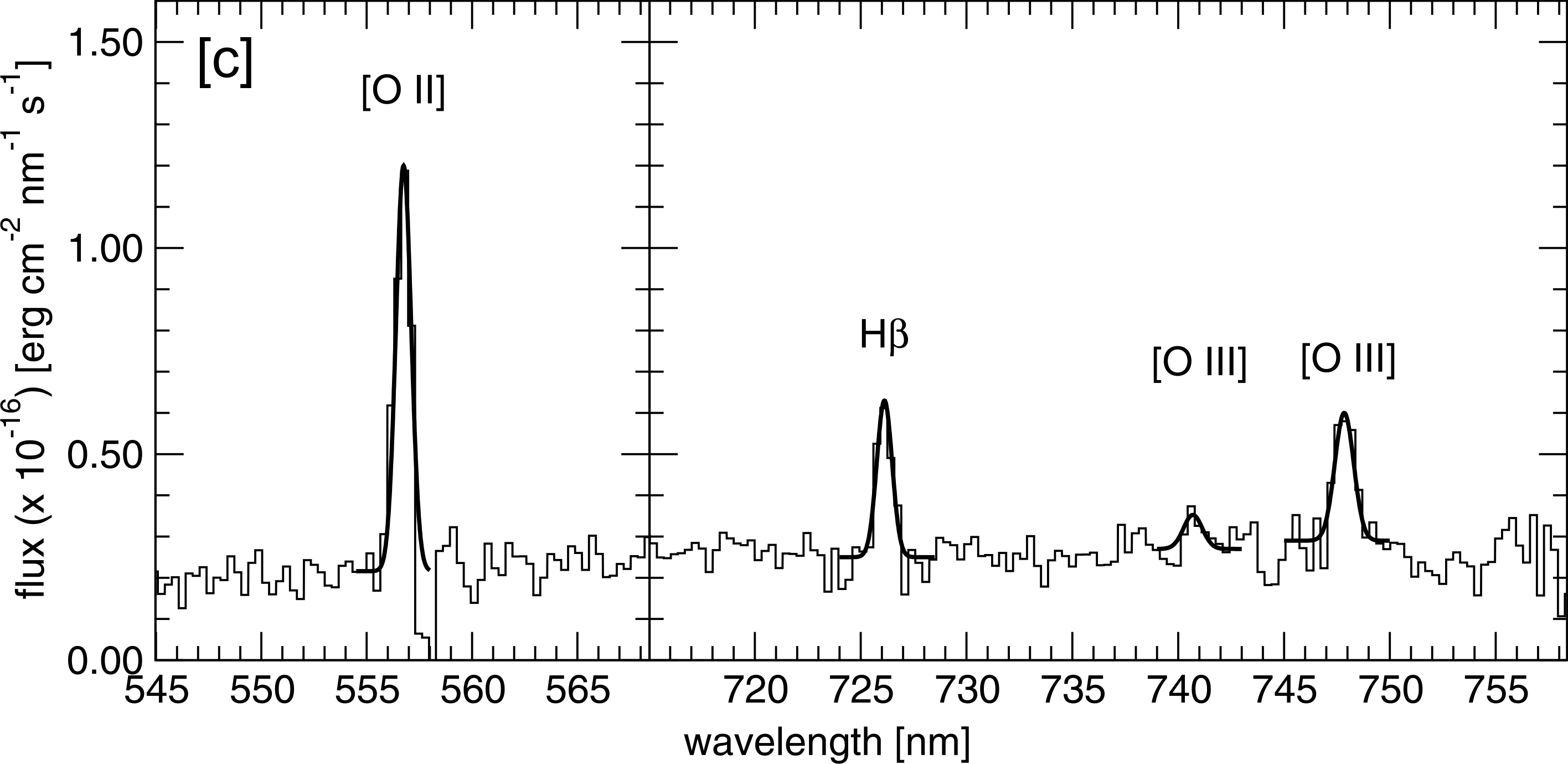}
      \caption{Galaxy \#3: The spectrum around [\ion{O}{ii}] (panel [a]) \ion{H}{$\beta$}+[\ion{O}{iii}]  (panel [b]) in the same intensity scale and size as Fig.~\ref{G1_image}. The spectral range in the upper panel is contaminated by remnants of the sky substraction of the green \ion{O}{i} airglow line at 557.7\,nm. For the integrated spectrum (panel [c]) the blue \ion{O}{iii} line was fitted assuming a fixed line intensity ratio of 1:3 between the two oxygen lines.
      }
         \label{G3_image}
   \end{figure}

\begin{figure}[!ht]
   \centering
   \includegraphics[width=65mm]{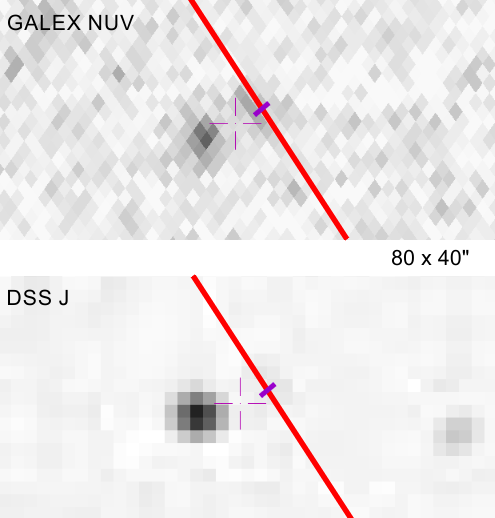}\\
        \caption{Position of galaxy \#3 (violet marker at the red FORS2 slit) on the blue DSS and the GALEX NUV band images. The cross marks the catalogue position of GALEXASC J215944.22-392017.3 .}
         \label{G3_id}
   \end{figure}

The coordinates in the slit of galaxy \#3 correspond fairly well to that of the source USNO$-$B1\,\,0506$-$0808786. Its $m_{\rm DSS\,\,J}$=21\fm1 also fits to the brightness of the continuum in the spectrum. The NED source GALEXASC J215944.22-392017.3 is 5\arcsec westwards half way to the about 15 magnitude 2MASS\,21594450-3920191 source, which certainly was not in the slit of our spectrum. Although the coordinates of the GALEX source and the uncertainty of only 2\farcs0 given in \citep{galex} do not fit well, we tend to believe that we identified this source since it was detected only in the NUV band of the GALEX as a 5\,$\sigma$ source (Fig.~\ref{G3_id}). Despite the redshift of nearly 0.5 it is the brightest of the sources in the optical continuum (Fig.~\ref{G3_image}).
Values of $\log(\ion{O}{iii}/\ion{H}{$\beta$}) = -0.09$ and $\log(\ion{O}{ii}/\ion{H}{$\beta$}) = +0.41$ (Table~\ref{G3_tab}) also indicate the galaxy being of {\sl composite} type in the scheme of \citet{line_ratio2}.

\begin{table}[!ht]
\caption{Summary of the observed parameters of galaxy \#3}
\label{G3_tab}
{\centering
\begin{tabular}{l c c c c}
\toprule
position$\!\!\!\!\!\!$ & \multicolumn{4}{l}{\qquad$\alpha_{\rm J2000} = \,\,\,\,21^{\rm h}59^{\rm h}44\fs0$}\\
\phantom{position}$\!\!\!\!\!\!$& \multicolumn{4}{l}{\qquad$\delta_{\rm J2000}\,=-39^{\rm o}20\arcmin15\arcsec$ }
\smallskip\\
ion  & $\lambda_{rest}$ & $\lambda_{obs}$ & $z$ & flux  \\
     & nm & nm &   & [*] \\
$[$\ion{O}{ii}$]$ & \,\,\,372.742$^{\rm a}$ & 556.706 & 0.4935 & \phantom{1}9.0 \\
\ion{H}{$\beta$}  & 486.135 & 726.104 & 0.4936 & \phantom{1}3.5\\
$[$\ion{O}{iii}$]$    & 495.891 & 740.753 & 0.4938 & ~--$^{\rm b}$ \\
                  & 500.684 & 747.879 & 0.4937 & \phantom{1}2.9\smallskip\\
redshift$\!\!\!\!\!\!$ & \multicolumn{4}{l}{\qquad$z_{\rm tot}=0.4937\pm0.0001$}
\smallskip\\
luminosity$\!\!\!\!\!\!\!\!\!\!\!\!\!$ & \multicolumn{4}{l}{\qquad$L_{{\rm H}\beta}\,\,=3.1\,\times\,10^{41} {\rm erg}\,{\rm cm}^{-2}\,{\rm s}^{-1}$}\\
 & \multicolumn{4}{l}{\qquad$L_{\ion{O}{iii}}=2.5\,\times\,10^{41} {\rm erg}\,{\rm cm}^{-2}\,{\rm s}^{-1}$}\\
 & \multicolumn{4}{l}{\qquad$L_{\ion{O}{ii}}\,=8.0\,\times\,10^{41} {\rm erg}\,{\rm cm}^{-2}\,{\rm s}^{-1}$}
\smallskip\\
\bottomrule
\end{tabular}}
$^{\rm a}\,\,\,\,${\small Calculated as mean of 372.603 and 372.881\,nm}\protect{\newline}
\phantom{$^{\rm a}\,\,\,\,$}{\small assuming about equal line strengths in the blend.}\protect{\newline}
$^{\rm b}\,\,\,\,${\small Only wavelength fitted - strength assumed 0.33 }\protect{\newline}
\phantom{$^{\rm b}\,\,\,\,$}{\small of the 500.684\,nm line.}
\end{table}

\subsection{Galaxy \#4 ($z=0.8668$): J215943.7$-$392012}
\begin{figure}[!ht]
   \centering
   \includegraphics[width=78mm]{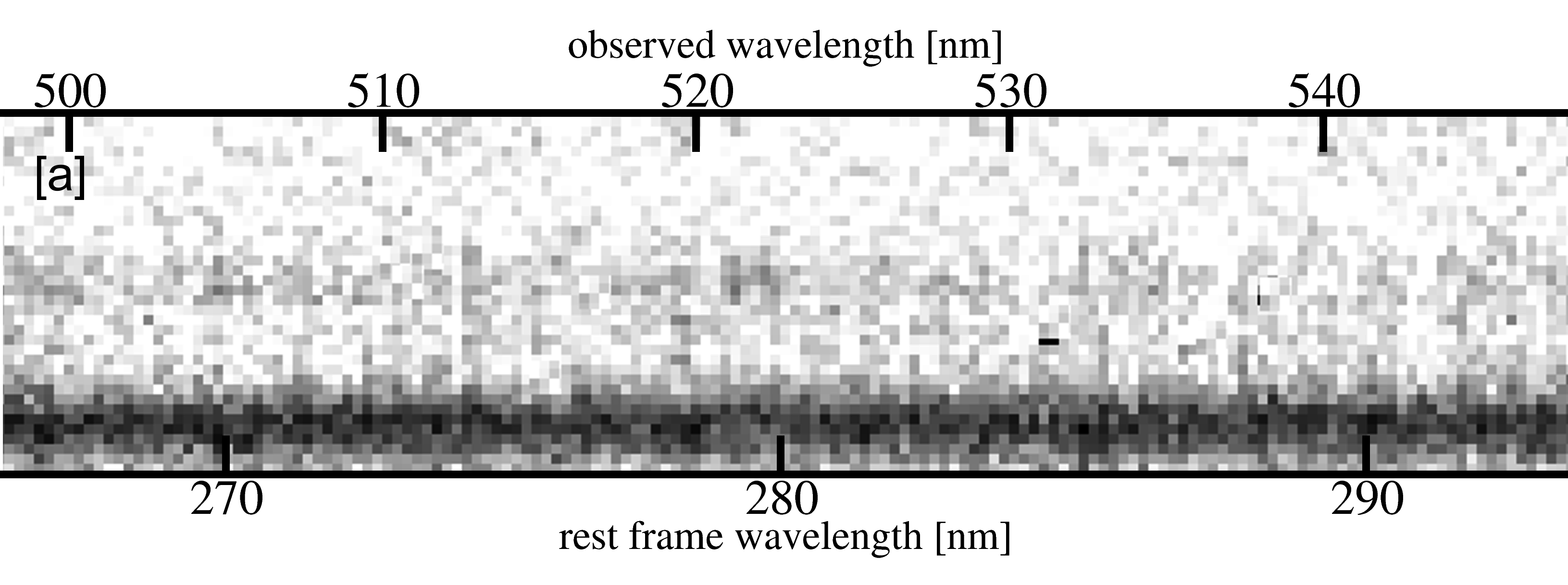}\\
   \smallskip
   \includegraphics[width=78mm]{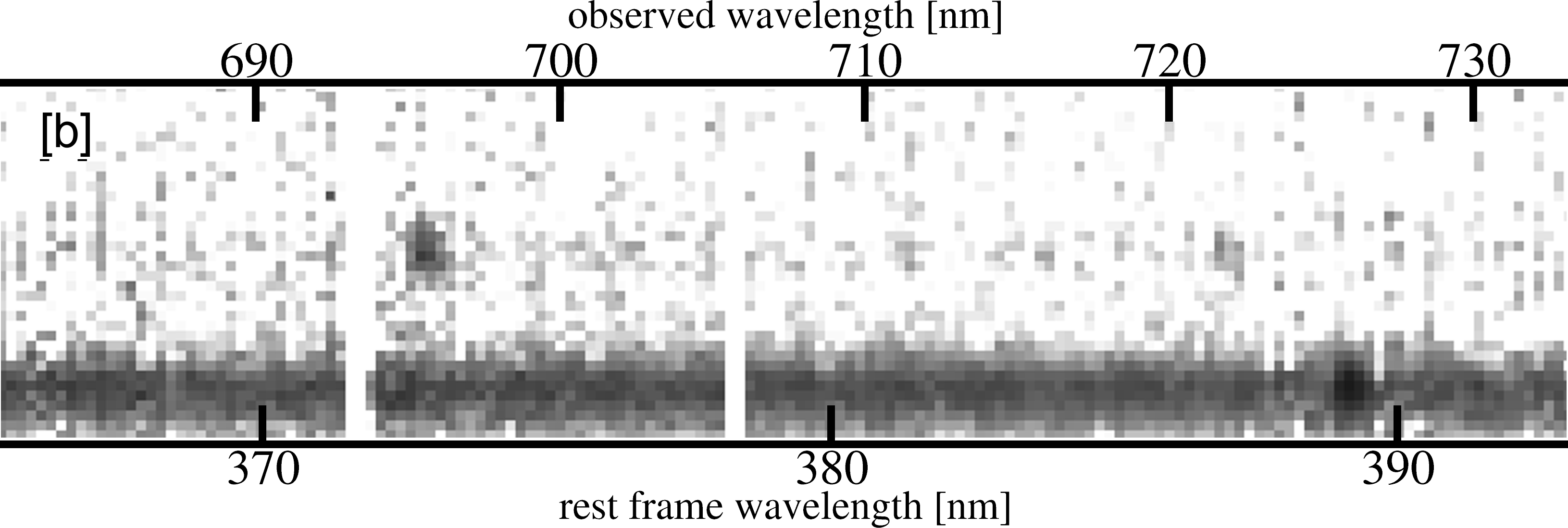}\\
   \smallskip
   \includegraphics[width=78mm]{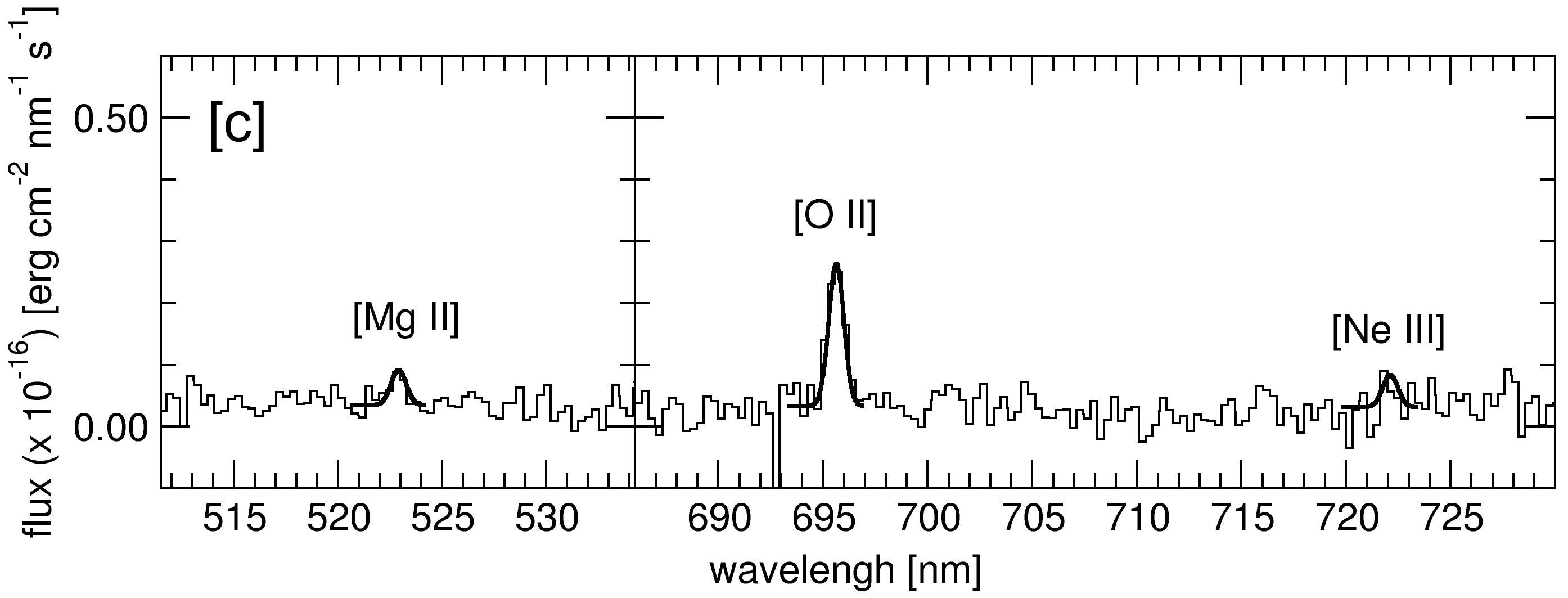}
            \caption{Galaxy \#4: The spectrum around [\ion{Mg}{ii}] (panel [a]) and around [\ion{O}{ii}] and [\ion{Ne}{iii}] (panel [b]). The bright target below is galaxy \#3 with the H$\beta$ at 726.10\,nm. Panel [c] shows the integrated spectrum of the galaxy and the best gaussian fits as overlay as given before.
      }
   \label{G4_image}
   \end{figure}

This target is only 3\farcs2 northeast of galaxy \#3. However, we found it to be the target with the highest redshift in our sample (Fig.\,\ref{G4_image}). It is the only target in our sample where the [\ion{Mg}{ii}] 279.8 nm UV line could be identified in the spectra, too (Table~\ref{G4_tab}). 

\begin{table}[!ht]
\caption{Summary of the observed parameters of galaxy \#4}
\label{G4_tab}
{\centering
\begin{tabular}{l c c c c}
\toprule
position$\!\!\!\!\!\!$ & \multicolumn{4}{l}{\qquad$\alpha_{\rm J2000} = \,\,\,\,21^{\rm h}59^{\rm h}43\fs7$}\\
\phantom{position}$\!\!\!\!\!\!$& \multicolumn{4}{l}{\qquad$\delta_{\rm J2000}\,=-39^{\rm o}20\arcmin12\arcsec$}
\smallskip\\
ion  & $\lambda_{rest}$ & $\lambda_{obs}$ & $z$ & flux  \\
     & nm & nm &   & [*] \\
$[$\ion{Mg}{ii}$]$ & 279.912 & 522.885 & 0.8677 & \phantom{1}0.5 \\
$[$\ion{O}{ii}$]$ & \,\,\,372.742$^{\rm a}$ & 695.621 & 0.8662 & \phantom{1}2.0 \\
\ion{Ne}{iii}  & 386.876 & 722.077 & 0.8664 & \phantom{1}0.5\smallskip\\
redshift$\!\!\!\!\!\!$ & \multicolumn{4}{l}{\qquad$z_{\rm tot}=0.8668\pm0.0008$}
\smallskip\\
luminosity$\!\!\!\!\!\!\!\!\!\!\!\!\!$ & \multicolumn{4}{l}{\qquad$L_{\ion{O}{ii}}\,=7.0\,\times\,10^{41} {\rm erg}\,{\rm cm}^{-2}\,{\rm s}^{-1}$}
\smallskip\\
\bottomrule
\end{tabular}}
$^{\rm a}\,\,\,\,${\small Calculated as mean of 372.603 and 372.881\,nm}\protect{\newline}
\phantom{$^{\rm a}\,\,\,\,$}{\small assuming about equal line strengths in the blend.}
\end{table}


\subsection{Galaxy \#5 ($z=0.7424$): J215935.61$-$2317.5}
\begin{figure}[!ht]
   \centering
   \includegraphics[width=78mm]{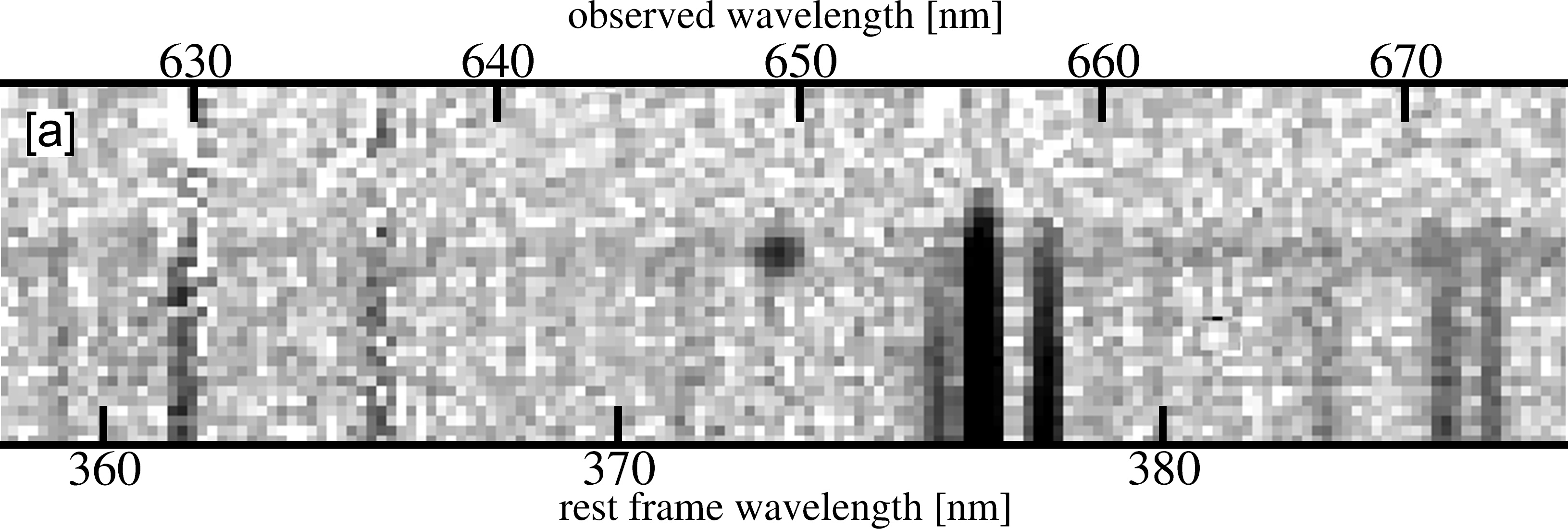}\\
   \smallskip
   \includegraphics[width=78mm]{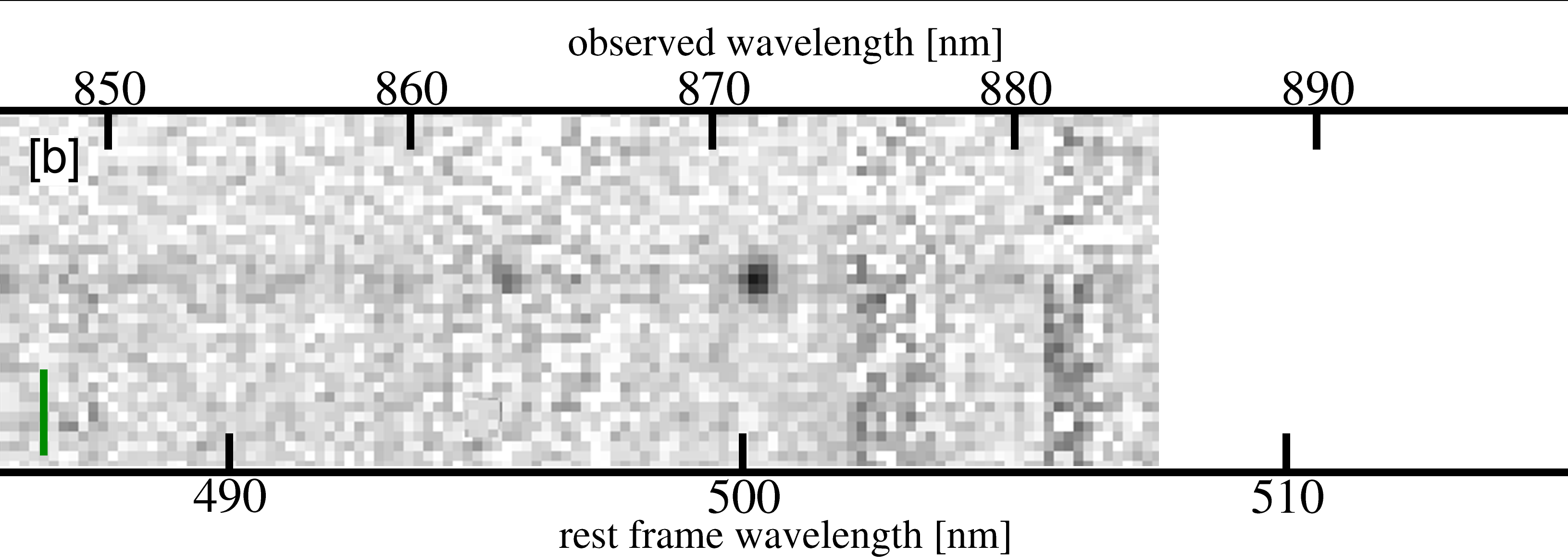}\\
   \smallskip
   \includegraphics[width=78mm]{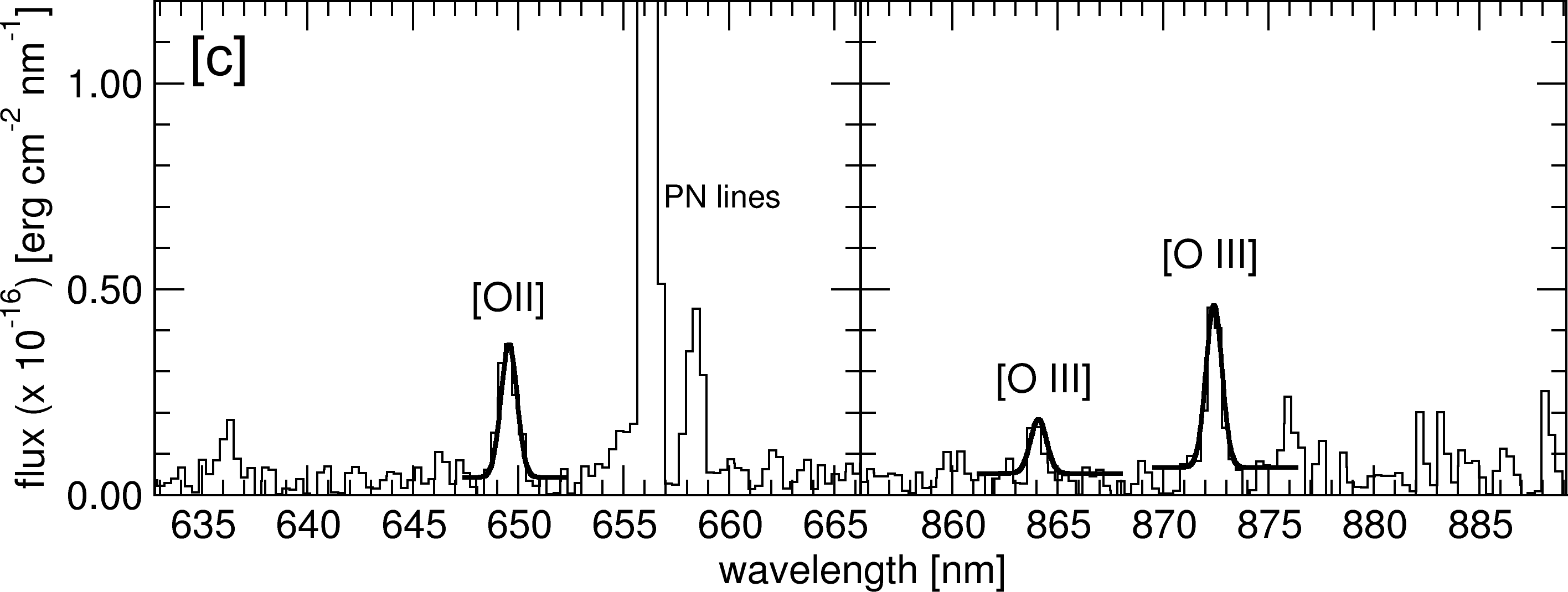}
      \caption{Galaxy \#5: Panel [a] shows strong remnants of the spatial inhomogeneous and thus only partially removed PN lines. The green mark in panel [b] shows the expected position of H$\beta$ which is lost due to bright OH telluric emission lines in that region. The right hand side is truncated due to the red end of our FORS2 setup.
      The integrated spectrum of the galaxy and the best gaussian fits as overlay are shown in panel [c]. }
         \label{G5_spec}
   \end{figure}

This target is very near to the PN center. Thus, the large variations of the PN emission lines do not allow a complete removal of them (Fig.~\ref{G5_spec}). Although the redshift should allow the detection of the [\ion{Mg}{ii}] 279.8 nm UV line, we could not detect it as its wavelength coincidences exactly with that of the \ion{H}{$\beta$} line of the PN. The \ion{H}{$\beta$} line of the galaxy is lost in telluric emission lines of the OH molecule (Table~\ref{G5_tab}). 

\begin{table}[!ht]
\caption{Summary of the observed parameters of galaxy \#5}
\label{G5_tab}
{\centering
\begin{tabular}{l c c c c}
\toprule
position$\!\!\!\!\!\!$ & \multicolumn{4}{l}{\qquad$\alpha_{\rm J2000} = \,\,\,\,21^{\rm h}59^{\rm h}35\fs61$}\\
\phantom{position}$\!\!\!\!\!\!$& \multicolumn{4}{l}{\qquad$\delta_{\rm J2000}\,=-39^{\rm o}23\arcmin17\farcs5$}
\smallskip\\
ion  & $\lambda_{rest}$ & $\lambda_{obs}$ & $z$ & flux  \\
     & nm & nm &   & [*] \\
$[$\ion{O}{ii}$]$ & \,\,\,372.742$^{\rm a}$ & 649.484 & 0.7424 & \phantom{1}3.1 \\
$[$\ion{O}{iii}$]$    & 495.891 & 863.948 & 0.7422 & \phantom{1}1.1 \\
                  & 500.684 & 872.414 & 0.7424 & \phantom{1}3.8 \smallskip\\
redshift$\!\!\!\!\!\!$ & \multicolumn{4}{l}{\qquad$z_{\rm tot}=0.7424\pm0.0001$}
\smallskip\\
luminosity$\!\!\!\!\!\!\!\!\!\!\!\!\!$ & \multicolumn{4}{l}{\qquad$L_{\ion{O}{ii}}\,=7.4\,\times\,10^{41} {\rm erg}\,{\rm cm}^{-2}\,{\rm s}^{-1}$}\\
&\multicolumn{4}{l}{\qquad$L_{\ion{O}{iii}}\,=9.0\,\times\,10^{41} {\rm erg}\,{\rm cm}^{-2}\,{\rm s}^{-1}$}
\smallskip\\
\bottomrule
\end{tabular}}
$^{\rm a}\,\,\,\,${\small Calculated as mean of 372.603 and 372.881\,nm}\protect{\newline}
\phantom{$^{\rm a}\,\,\,\,$}{\small assuming about equal line strengths in the blend.}
\end{table}

\subsection{Galaxy \#6 ($z=0.3137$): J215942.4$-$392530}
\begin{figure}[!ht]
   \centering
   \includegraphics[width=78mm]{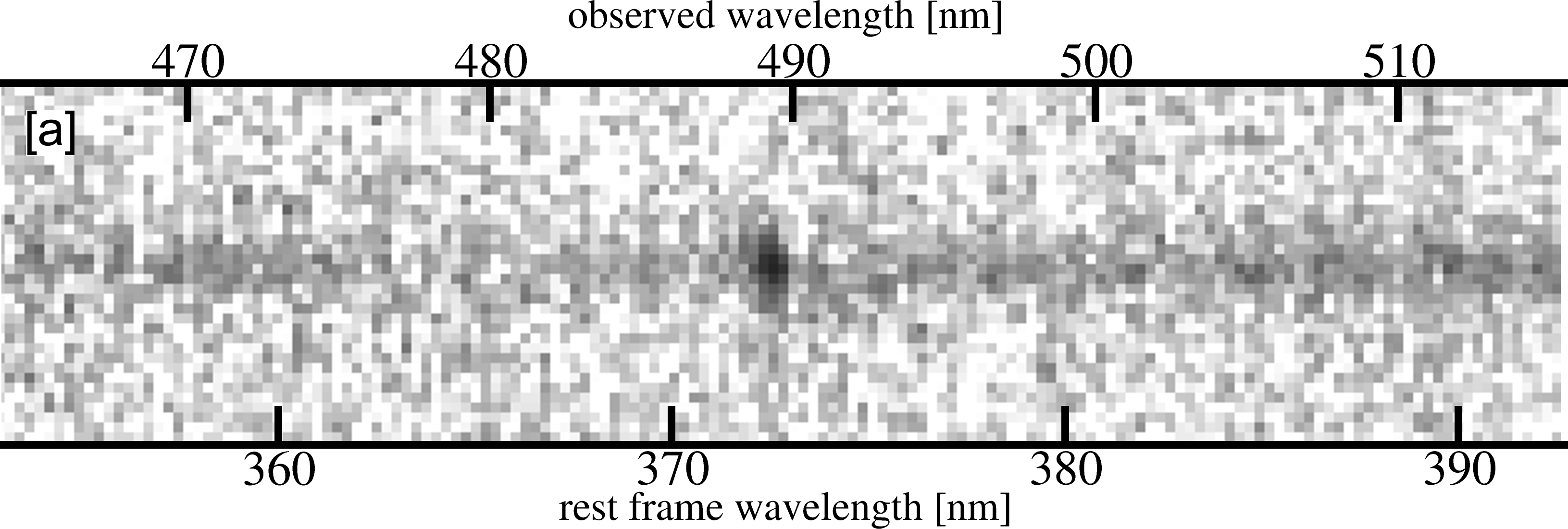}\\
   \smallskip
   \includegraphics[width=78mm]{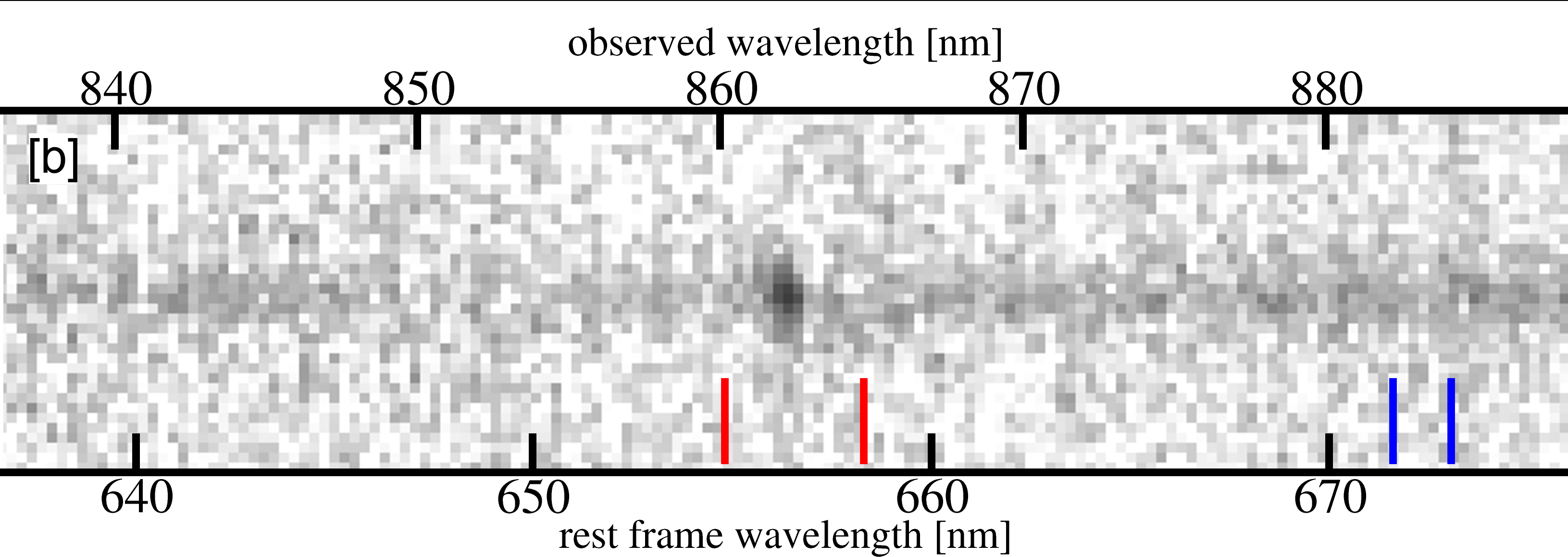}\\
   \smallskip
   \includegraphics[width=78mm]{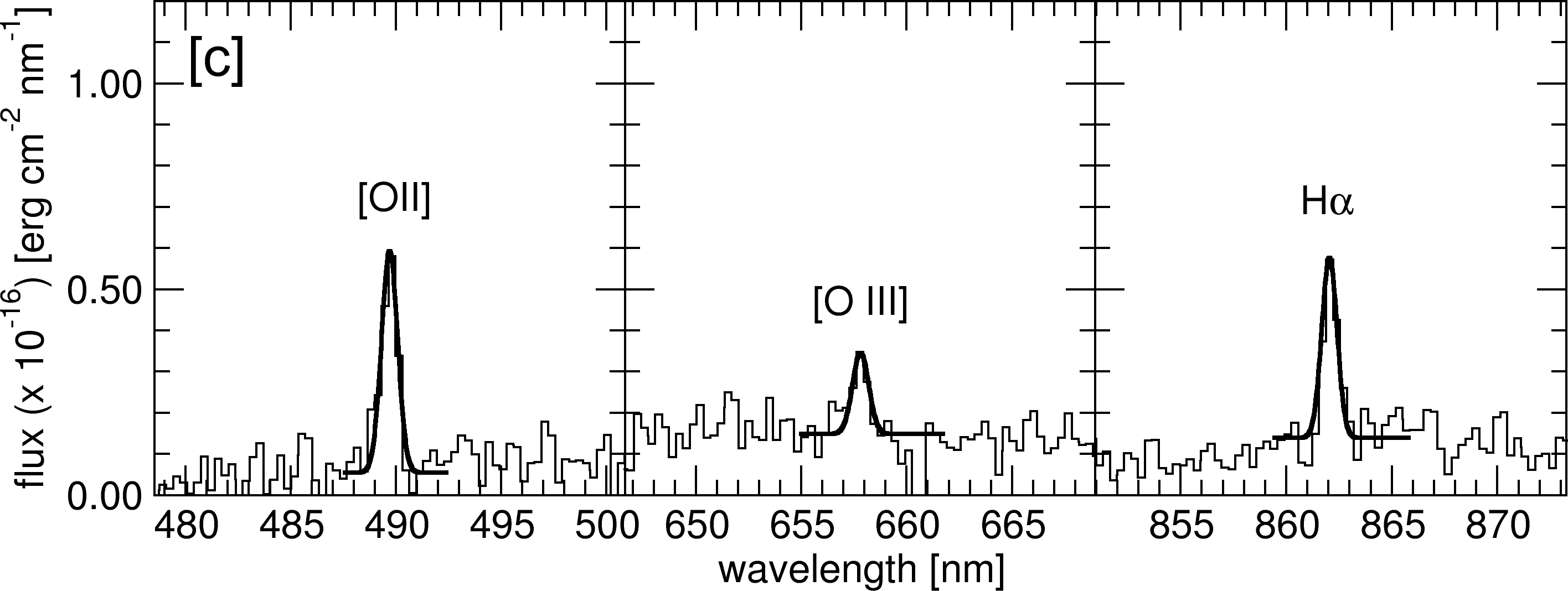}
      \caption{Galaxy \#6. Panel [a] shows the spectrum around [\ion{O}{ii}]. As in case of galaxy \#1, the position of the nitrogen and sulphur lines are given in panel [b]. However, due to strong telluric OH molecular lines in that region a detection was not possible. Panel [c] again shows the integrated spectrum.  }
         \label{G6_spec}
   \end{figure}

\begin{figure}[!ht]
   \centering
   \includegraphics[width=65mm]{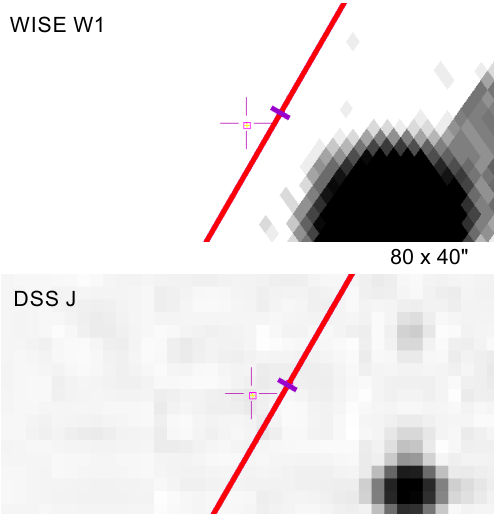}\\
        \caption{Position of galaxy \#6 (violet marker at the red FORS2 slit) on the blue DSS and the WISE W1 band images. The cross marks the catalogue position of WISE J214941.03-392547.1 .}
         \label{G6_id}
   \end{figure}

This target is located behind the thin faint PN halo. Although the wavelength range covers the spectral region around \ion{H}{$\alpha$}, the [\ion{N}{ii}] and [\ion{S}{ii}] lines were not visible due to strong OH line contamination. (Fig.~\ref{G6_spec}, Table~\ref{G5_tab}).
The positional uncertainties of 0\farcs6 given in the catalog of WISE J215942.81$-$392533.2 do not fit to the 3\farcs5 distance to our target. Nevertheless, we believe these objects to be identical. Visual inspection of the WISE image \citep{AllWISE} at CDS/Aladin \citep{aladin} show a very weak source at the detection limit and the bright stellar source WISE J215941.03$-$392547.1 nearby has a FWHM of 8\farcs2 on the image (Fig.~\ref{G6_id}). Thus the realistic positional error in the WISE data should be higher.

\begin{table}[!ht]
\caption{Summary of the observed parameters of galaxy \#6}
\label{G6_table}
{\centering
\begin{tabular}{l c c c c}
\toprule
position$\!\!\!\!\!\!$ & \multicolumn{4}{l}{\qquad$\alpha_{\rm J2000} = \,\,\,\,21^{\rm h}59^{\rm h}42\fs4$}\\
\phantom{position}$\!\!\!\!\!\!$& \multicolumn{4}{l}{\qquad$\delta_{\rm J2000}\,=-39^{\rm o}25\arcmin30\arcsec$}
\smallskip\\
ion  & $\lambda_{rest}$ & $\lambda_{obs}$ & $z$ & flux  \\
     & nm & nm &   & [*] \\
$[$\ion{O}{ii}$]$ & \,\,\,372.742$^{\rm a}$ & 489.714 & 0.3138 & \phantom{1}4.6 \\
$[$\ion{O}{iii}$]$    & 495.891 & 651.602 & \,\,\,0.3140$^{\rm b}$ & \phantom{1}0.6 \\
                  & 500.684 & 657.786 & 0.3137 & \phantom{1}1.4 \\
\ion{H}{$\alpha$}  & 386.876 & 862.125 & 0.3137 & \phantom{1}4.0\smallskip\\
redshift$\!\!\!\!\!\!$ & \multicolumn{4}{l}{\qquad$z_{\rm tot}=0.3137\pm0.0001$}
\smallskip\\
luminosity$\!\!\!\!\!\!\!\!\!\!\!\!\!$ & \multicolumn{4}{l}{\qquad$L_{\ion{O}{ii}}\,=4.6\,\times\,10^{41} {\rm erg}\,{\rm cm}^{-2}\,{\rm s}^{-1}$}\\
&\multicolumn{4}{l}{\qquad$L_{\ion{O}{iii}}\,=4.3\,\times\,10^{40} {\rm erg}\,{\rm cm}^{-2}\,{\rm s}^{-1}$}\\
&\multicolumn{4}{l}{\qquad$L_{\ion{H}{$\alpha$}}\,=1.2\,\times\,10^{41} {\rm erg}\,{\rm cm}^{-2}\,{\rm s}^{-1}$}\smallskip\\
\bottomrule
\end{tabular}}
$^{\rm a}\,\,\,\,${\small Calculated as mean of 372.603 and 372.881\,nm}\protect{\newline}
\phantom{$^{\rm a}\,\,\,\,$}{\small assuming about equal line strengths in the blend.}\protect{\newline}
$^{\rm b}\,\,\,\,${\small Not included to calculate value of $z_{\rm tot}$ }\protect{\newline}
\phantom{$^{\rm b}\,\,\,\,$}{\small due line weakness.}
\end{table}

\section{Diagnostic diagrams}

Although the sample is small and only a few line ratios are completely given one can get a first guess about these galaxies in the classical diagnostic diagrams following the BPT scheme \citep{BPT}. We applied the same analysis as proposed by \citet{line_ratio3}, \citet{line_ratio1} and \citet{line_ratio2}. As our sample covers a redshift domain from $0.2<z<0.9$ we selected the VIMOS VLT Deep Survey (VVDS) data set of \citet{VVDS}, studying a large data set with the same redshift domain.

\begin{figure}[!ht]
   \centering
   \includegraphics[width=78mm]{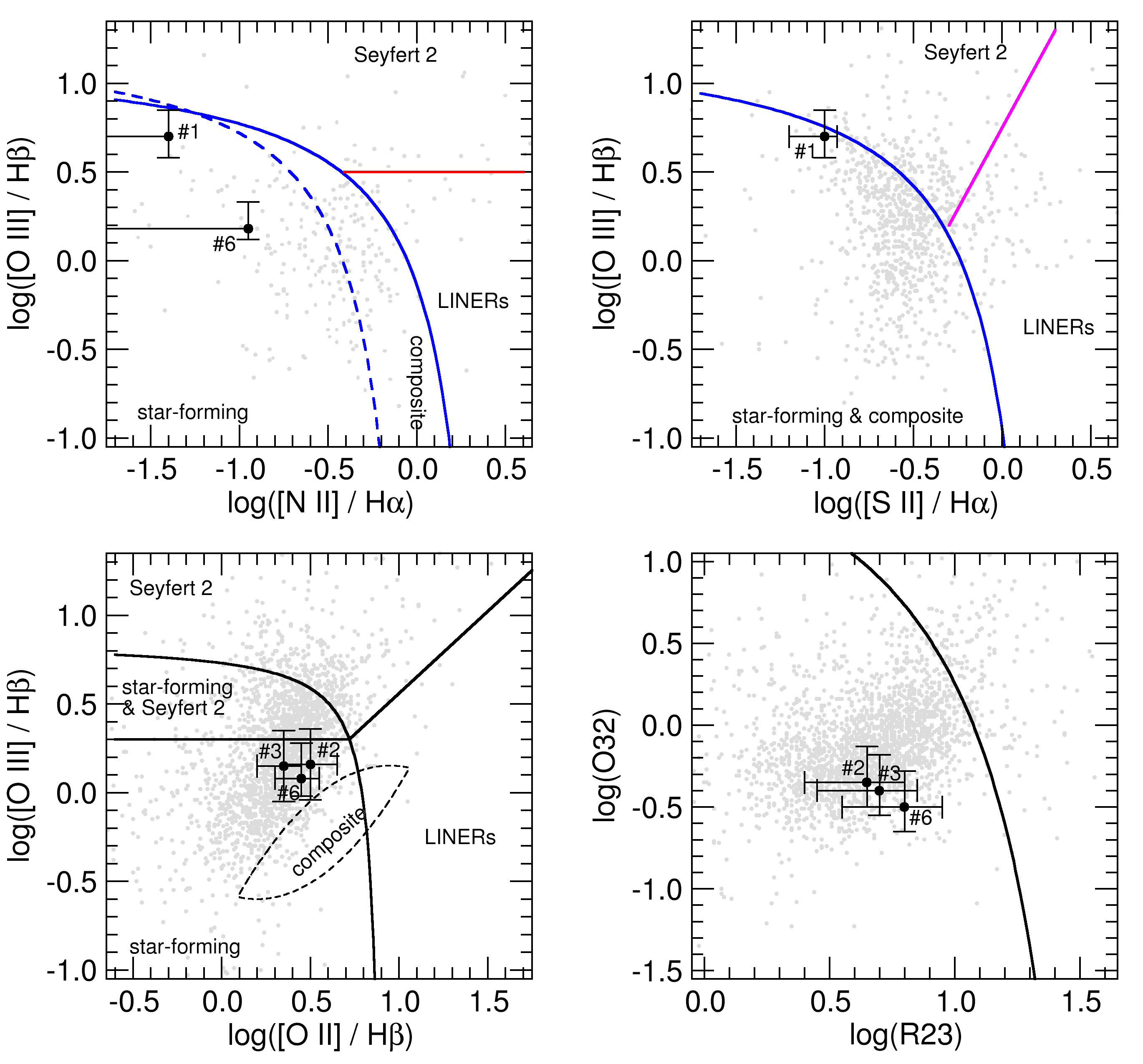}\\
        \caption{Diagnostic diagrams following the classical BPT in the version of \citet{line_ratio1}. The delimiter lines between the different galaxy types were defined by \citet{VeOs87}, \citet{Kewley01}, \citet{Kauffmann03} and \citet{line_ratio3}. The underlying grey symbols are the galaxies of the VVDS. Open one sided error bars denote upper limits.}
         \label{fig_BPT}
   \end{figure}

Galaxies \#1 and \#6 are the only ones having redshifts allowing to investigate the region around \ion{H}{$\alpha$}. Although the \ion{S}{ii} and \ion{N}{ii} lines are limits or marginally detected , giving large error bars. In case of \#6 the \ion{H}{$\beta$}) was estimates from the theoretical \ion{H}{$\alpha$}/\ion{H}{$\beta$} ratio and the error bars assume the unknown extinction to be from $0<A_{\rm V}<3^{\rm m}$.
For galaxies \#2 and \#3 we can use the blue line ratios directly and for \#6 with the same access as above too (Fig.~\ref{fig_BPT}). For \#4 and \#5 we do not have any complete pair of line ratios.
All galaxies seem to belong to the family of normal star forming galaxies, although \#1 is very near to the border to the Seyfert~2 galaxies.
\section{Summary and conclusion}
We reported on the discovery of six emission line galaxies within only two long-slit observation taken with the FORS2 spectrograph. The large spread of redshifts show, that these galaxies not are the result of a single bound cluster structure. Although upcoming large surveys on ELGs will give a very high statistics in nearby future, the serendipitous discoveries reported here show the potential of a survey based on the FORS2 ESO archive. Since we detected our sample in only two slit pointings covering a sky area of only 570 arcsec$^2$, we expect the number of potential further candidates to be very large, in particular since FORS2 is in operation for more than a decade as one of the first-light VLT instruments. FORS2 is also one of the work horses at the VLT, and the spectrograph with the largest duty cycle without major changes on 8m class telescopes, leading to frequent scheduling and a high pointing coverage on the entire sky. Conducting a systematic serendipity survey based on FORS archival data will therefore lead to a large sample of ELGs completely unbiased by any selection criteria based on spatial distributions.
That would be, as to our knowledge, the very first attempt to use archival observations on optical slit spectroscopy to obtain a {\sl Serendipity Survey}. Although the work of \citet{1995AJ....110..982T} is called {\sl Serendipitous Long-Slit Surveys for Primeval Galaxies} it is only comparable in technique. It was obtained form dedicated pointing observations for this very extragalactic survey and not as byproduct of other observations and thus is restricted to a very limited deep field.
Looking to the ESO archive we found about 20\,000 pointings at galactic latitudes above 20$^{\rm o}$~and appropriate exposure times of $\ge$15 minutes.
Moreover using this approach we do not introduce a photometric pre-selection, as e.g. \citet{eBOSS},
leading to an unbiased sample. The size of the telescope, being much larger than the typical 2-4m class survey telescopes, and the exposures often going to one hour and more, will pick also galaxies with fairly small emission line contrast on top of their continuum.

As the publication of the first 20\% of the MUSE-Wide survey \citep{MUSE} shows, this instrument will add an additional possibility soon, when more data gets public. They detected in the dedicated pointings 831 emission line galaxies in 22 arcmin$^2$ on the sky with 1 hour exposures each. Their survey is 140 times larger and the exposure time was about a factor of 2 larger. Downsizing to our test area of 0.16 arcmin$^2$, the number of 5.9 expected  detections is pretty much the same as we got. A future extension to MUSE thus seems to be an attractive option too, although the required data handling will be more sophisticated.

\begin{acknowledgements}
\noindent The study is based on observations made with ESO Telescopes at the La Silla Paranal Observatory under programme ID 098.D-0332.\newline
This research has made use of the SIMBAD database,
operated at CDS, Strasbourg, France \citep{simbad}, the NASA/IPAC Extragalactic Database (NED) which is operated by the Jet Propulsion Laboratory, California Institute of Technology, under contract with the National Aeronautics and Space Administration and has made use of "Aladin sky atlas" developed at CDS, Strasbourg Observatory, France \citep{aladin}.
Daniela Barria and this publication was financed by the ALMA--CONICYT Fund, allocated to the project N$^{\rm o}\,31150001$ and Wolfgang Kausch is supported by the project IS538003 (Hochschulraumstrukturmittel) provided by the Austrian Ministry for Research, Investigation and Economy (BM:wfw).
\end{acknowledgements}


\relax
\end{document}